\documentclass[aps,prx,twocolumn,balance,superscriptaddress,floats,longbibliography]{revtex4-1}

\pdfoutput=1
\usepackage[a4paper, total={7.2in, 10in}]{geometry}
\usepackage{latexsym}
\usepackage{dcolumn}
\usepackage{amssymb,amsmath}
\usepackage{epsf}
\usepackage{float}
\usepackage{hyperref}
\usepackage{mathrsfs}

\usepackage[pdftex]{graphicx}
\usepackage{epstopdf}
\usepackage{upgreek}

\usepackage{xcolor}
\usepackage{soul}

\begin{document}

%%\preprint{APS/123-QED}

\title{Resonantly enhanced nonreciprocal silicon Brillouin amplifier}

\author{Nils T. Otterstrom}
\email{nils.otterstrom@yale.edu}
\affiliation{Department of Applied Physics, Yale University, New Haven, CT 06520 USA.}
\author{Eric A. Kittlaus}
\affiliation{Department of Applied Physics, Yale University, New Haven, CT 06520 USA.}
\author{Shai Gertler}
\affiliation{Department of Applied Physics, Yale University, New Haven, CT 06520 USA.}
\author{Ryan O. Behunin}
\affiliation{Department of Physics and Astronomy, Northern Arizona University, Flagstaff, AZ 86001 USA.}
\author{Anthony L. Lentine}
\affiliation{Applied Photonic Microsystems, Sandia National Laboratories, Albuquerque, New Mexico 87185, USA}
\author{Peter T. Rakich}
\email{peter.rakich@yale.edu}
\affiliation{Department of Applied Physics, Yale University, New Haven, CT 06520 USA.}

%\makeatother

\date{\today}

\begin{abstract}

The ability to amplify light within silicon waveguides is central to the development of high-performance silicon photonic device technologies. To this end, the large optical nonlinearities made possible through stimulated Brillouin scattering offer a promising avenue for power-efficient all-silicon amplifiers, with recent demonstrations producing several dB of net amplification. However, scaling the degree of amplification to technologically compelling levels ($>10$ dB), necessary for everything from filtering to small signal detection, remains an important goal.  Here, we significantly enhance the Brillouin amplification process by harnessing an inter-modal Brillouin interaction within a multi-spatial-mode silicon racetrack resonator.  Using this approach, we demonstrate more than 20 dB of net Brillouin amplification in silicon, advancing state-of-the-art performance by a factor of 30. This degree of amplification is achieved with modest ($\sim15$ mW) continuous-wave pump powers and produces low out-of-band noise. Moreover, we show that this same system behaves as a unidirectional amplifier, providing more than 28 dB of optical nonreciprocity without insertion loss in an all-silicon platform. Building on these results, this device concept opens the door to new types of all-silicon injection-locked Brillouin lasers, high-performance photonic filters, and waveguide-compatible distributed optomechanical phenomena.

\end{abstract}

\maketitle

\section{Introduction}

High-performance optical amplification is an essential functionality in integrated photonic circuits. Within the context of silicon photonics, however, strategies for robust integrated amplifiers have faced significant challenges that stem from silicon’s indirect bandgap and high levels of nonlinear loss \cite{leuthold2010nonlinear,liang2010recent}. To date, silicon amplifier technologies have relied on either hybrid integration strategies \cite{park2007hybrid,heck2011hybrid,keyvaninia2012highly,agazzi2010monolithic} or nonlinear optical interactions \cite{leuthold2010nonlinear} such as Raman \cite{claps2003observation,boyraz2004demonstration,liang2004efficient,liu2004net} or Kerr effects \cite{foster2006broad,liu2010mid,kuyken201150}. While Raman and Kerr interactions have been used to produce net amplification using pulsed optical pumping \cite{claps2003observation,boyraz2004demonstration,liang2004efficient,liu2004net,foster2006broad}, active electrical removal of free carriers \cite{rong2005continuous}, or large mid-IR pump powers \cite{liu2010mid,kuyken201150}, it remains nontrivial to achieve large degrees of power-efficient optical amplification due to the competition between gain and nonlinear absorption in silicon.

Recently, nonlinear light-sound coupling known as stimulated Brillouin scattering has emerged as a promising mechanism for optical amplification in silicon, with dynamics and performance that can be customized through structural control \cite{rakich2012giant,shin2013tailorable}.  Once entirely absent from silicon photonics, these Brillouin interactions have emerged as one of the strongest and most tailorable nonlinearities in silicon \cite{shin2013tailorable,van2015interaction}, opening the door to net optical amplification \cite{van2015net,kittlaus2016large,kittlaus2017chip} and Brillouin lasing in silicon photonic circuits \cite{otterstrom2018silicon}. While recent demonstrations have achieved 2-5 dB of amplification \cite{kittlaus2016large,kittlaus2017chip}, scaling the degree of amplification to levels necessary for high-fidelity filtering \cite{tanemura2002narrowband,pant2014chip,souidi2016low,choudhary2017advanced} and small-signal detection schemes \cite{kozlovskiui2006detection} remains a nontrivial challenge \cite{wolff2015power}. 

\begin{figure*}[ht]
\centering%\vspace{-10pt}
\includegraphics[width=.9\linewidth]{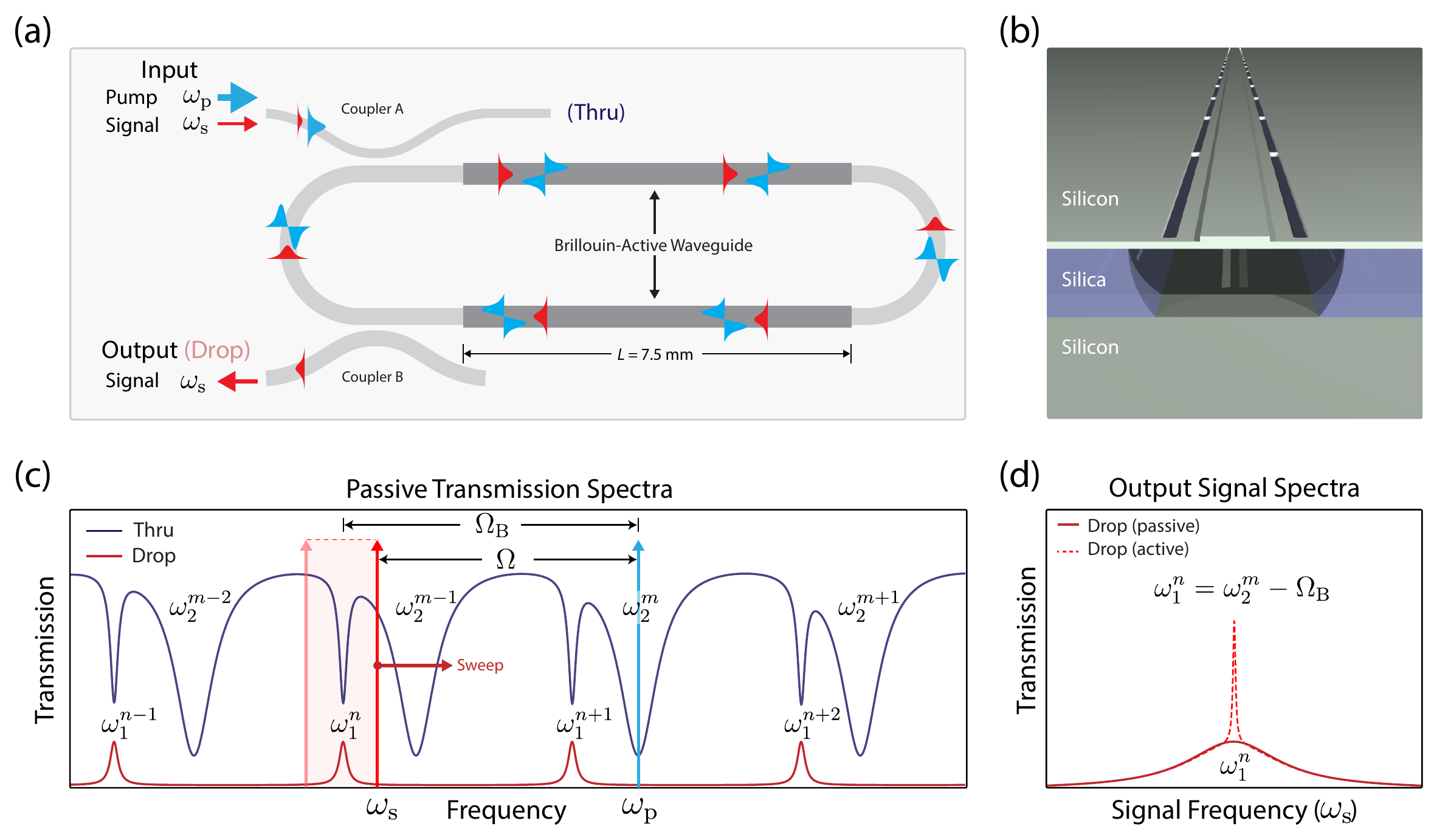}
\caption{(a) Resonantly enhanced Brillouin amplifier device concept and operation scheme. The amplifier is composed of a multi-spatial-mode racetrack resonator with two Brillouin-active regions. Using the frequency selectivity of the cavity, pump ($\omega_{\rm p}$) and and signal waves ($\omega_{\rm s}$) are coupled into the antisymmetric and symmetric cavity modes, respectively, via a multimode coupler. As the pump and signal waves traverse the Brillouin-active segments, the pump wave resonantly amplifies the signal wave through stimulated inter-modal Brillouin scattering. The signal wave exits the system through a mode-selective coupler (drop port), which is designed to couple strongly to the symmetric mode and weakly to the antisymmetric mode. (b) Schematic illustrating the cross-sectional geometry of the Brillouin-active regions. This suspended multimode silicon waveguide supports two transverse electric (TE)-like optical spatial modes and a 6-GHz antisymmetric Lamb-like elastic wave, which mediates inter-modal Brillouin amplification. (c) Idealized optical transmission spectra at the thru and drop ports. Coupling into the racetrack resonator via a multimode coupler yields a characteristic multimode transmission spectrum at the thru port, with broad (centered at $\omega^{m}_{2}$) and narrow (centered at $\omega^{n}_{1}$) resonances corresponding to the antisymmetric and symmetric optical spatial modes, respectively. The mode-selective drop port is designed to couple out only the symmetric cavity modes. Resonantly-enhanced Brillouin amplification measurements are performed by coupling the pump wave ($\omega_{\rm p}$) to an antisymmetric cavity mode ($\omega^{m}_{2}$) and sweeping the signal wave ($\omega_{\rm s}$) through a symmetric cavity mode ($\omega^{n}_{1}$) that is red-shifted from  by the Brillouin frequency ($\Omega_{\rm B}$). (d) Zoomed-in transmission spectrum for the signal wave exiting the drop port when $\omega_{\rm s}\approx \omega^{n}_{1}$ with (active) and without (passive) the Brillouin gain supplied by the pump wave.    }
\label{fig:deviceconcept}
\end{figure*}

In this letter, we demonstrate record-high Brillouin gain and amplification in a silicon waveguide through a resonantly enhanced Brillouin interaction. This all-silicon amplifier system is based on a stimulated inter-modal Brillouin scattering process, in which a traveling elastic wave mediates nonlinear energy transfer between light waves propagating in distinct optical spatial modes \cite{kittlaus2017chip}. Building on existing device concepts for silicon Brillouin lasers \cite{otterstrom2018silicon}, we harness and dramatically enhance this stimulated Brillouin process using a multi-spatial-mode racetrack resonator system that is interfaced with mode-specific couplers to allow signal light to be amplified as it is transmitted through the system. We use this device to realize 30 dB of Brillouin gain, corresponding to over 20 dB of net Brillouin amplification. These results represent a $>30$-fold improvement beyond state-of-the-art Brillouin amplification in silicon \cite{kittlaus2016large}. Leveraging the unidirectional amplification produced by this phase-matched process, we also use this system to demonstrate more than 28 dB of nonreciprocal contrast between forward- and backward-propagating waves. This scheme provides robust optical nonreciprocity without insertion loss.  Beyond the results presented here, this device concept opens the door to chip-integrated injection-locked Brillouin lasers, microwave-photonic filtering techniques, and distributed optomechanical phenomena.

\begin{figure*}[ht]
\centering%\vspace{-10pt}
\includegraphics[width=\linewidth]{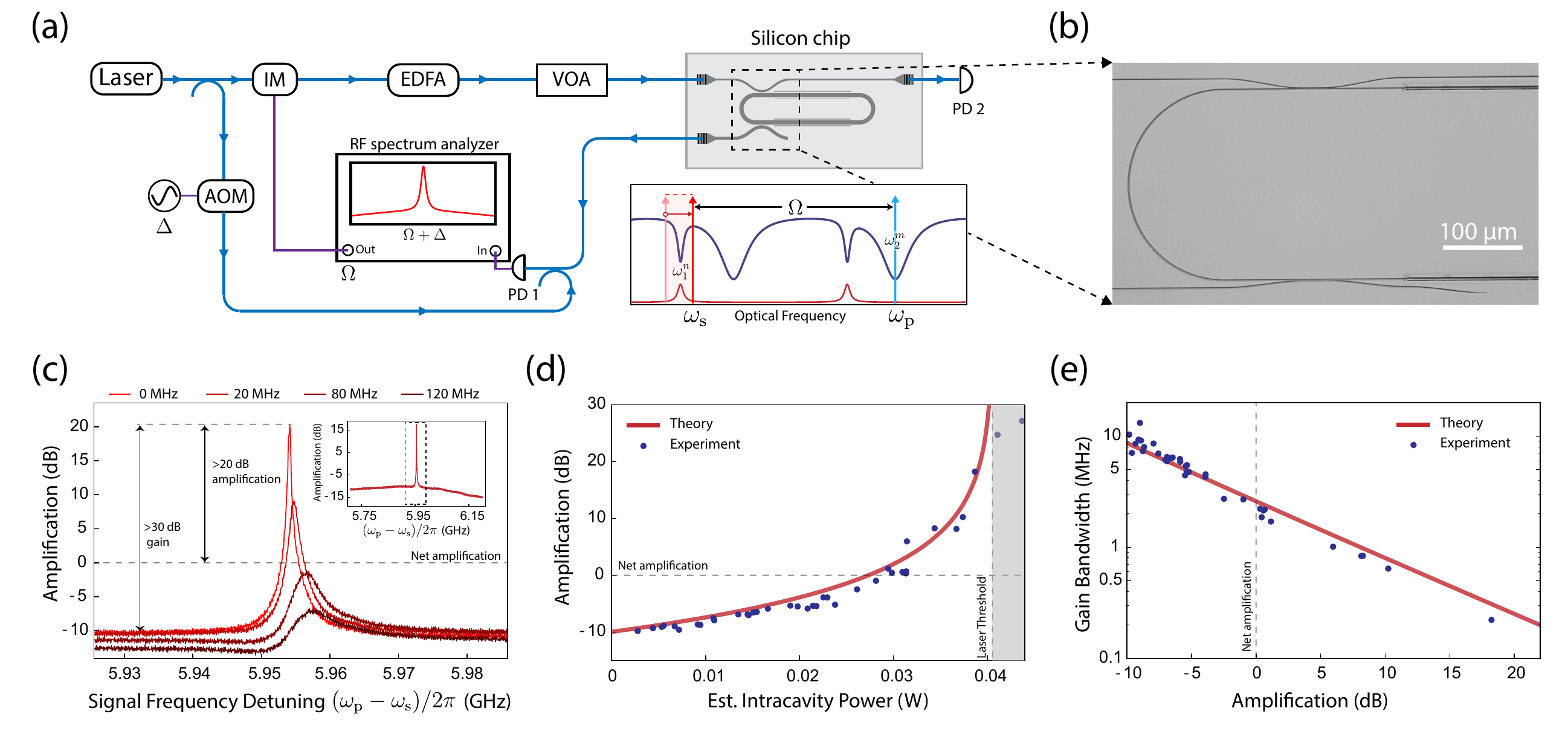}
\caption{(a) Diagram of the experimental apparatus used to characterize the resonantly enhanced Brillouin amplifier.  Laser light is split along two paths. One path is used to synthesize an optical local oscillator (LO) using an acousto-optic modulator (AOM), which blueshifts the light by $\Delta= 2 \pi \times 44$ MHz. The other arm synthesizes pump and signal waves with the desired frequency detuning ($\Omega=\omega_{\rm p}-\omega_{\rm s}$) and powers using an intensity modulator (IM), erbium-doped-fiber amplifer (EDFA), and variable optical attentuator (VOA); the light is subsequently coupled on-chip for nonlinear amplification measurements. After passing through the device, the signal wave is coupled through the drop port and off-chip, where it is combined with the blueshifted LO and measured on a high-speed photodector (PD 1). The RF spectrum analyzer sweeps the detuning ($\Omega$) and measures the microwave power at ($\Omega+\Delta$), permitting single-sideband measurements of $\omega_{\rm s}=\omega_{\rm p}-\Omega$ (without crosstalk from light at $\omega_{\rm p}+\Omega$). (b) Optical micrograph (in gray scale) showing a top-down view of part of the device. (c) Gain spectra as a function of signal-wave detuning around the Brillouin resonance, showing more than 30 dB of gain and 20 dB of net amplification.  Each trace represents a different estimated detuning of the optical cavity mode relative to the Brillouin freqeuncy (see zoomed-out inset). Large optical cavity detunings relative to the Brillouin resonance diminish the degree of amplification and result in characteristic asymmetric line-shapes. (d) Measured and theoretical signal-wave amplification produced over a range of intracavity powers. As the pump power approaches the laser threshold power, the resonantly enhanced Brillouin amplification increases dramatically. Data is compiled from a series of power, microwave-frequency detuning, and wavelength sweeps (for more details see Appendix C2). (e) Linewidth narrowing of the gain bandwidth as a function of signal wave amplification. }
\label{fig:data}
\end{figure*}

\section{Results}
We use a resonant optical configuration to produce greatly enhanced inter-modal Brillouin amplification and optical nonreciprocity in an all-silicon structure.  This strategy allows us to transform the otherwise modest amplification ($\sim2 $ dB) possible in a linear geometry (using a waveguide of the same design; see Ref. \cite{kittlaus2017chip}), into more than 20 dB of net amplification.  We achieve this significant enhancement in performance by leveraging the resonator geometry diagrammed in Fig. \ref{fig:deviceconcept}a that builds upon the laser structure described in Ref. \cite{otterstrom2018silicon}.  In this type of configuration, resonant transmission through the system becomes lossless (i.e., approaches unity) if the internal gain produced by stimulated inter-modal Brillouin scattering balances the internal losses of the resonator.  In the case when the gain exceeds the internal losses of the resonator, but does not exceed the total loss (i.e., internal + external), the system can yield greater-than-unity transmission without producing self-oscillation (i.e., below the laser threshold). In the limit when the gain approaches the total loss of the system, the degree of amplification can become arbitrarily large---in principle, limited only by gain depletion. Furthermore, we show that due to the phase-matching requirements of the stimulated inter-modal Brillouin process, this resonantly enhanced amplification is unidirectional, yielding large degrees of optical nonreciprocity.

The resonantly enhanced Brillouin amplifier consists of a 15 mm-long racetrack cavity that possesses two Brillouin-active regions (Fig. \ref{fig:deviceconcept}a). The device is fabricated from a single-crystal SOI wafer using a hybrid CMOS-MEMS process (see Appendix C3).  The racetrack cavity is formed from a multimode silicon ridge waveguide that supports low-loss guidance of transverse electric (TE)-like symmetric and antisymmetric optical spatial modes. In the Brillouin-active regions, this multimode optical waveguide is suspended to provide acoustic guidance for the 6-GHz traveling elastic wave that mediates strong inter-modal Brillouin coupling (see Fig. \ref{fig:deviceconcept}b; for waveguide dimensions, see Appendix C3). Within the resonator, the symmetric and antisymmetric optical spatial modes form two distinct sets of resonances.  The symmetric spatial mode produces a set of high-Q factor ($Q_1=10^6$) cavity modes centered at frequencies $\{\omega_1^{n}\}$ while the antisymmetric spatial mode produces a set of cavity modes having lower Q factors ($Q_2=2\times10^5$) at frequencies $\{\omega_2^{m}\}$.

To access these cavity modes, the resonator is interfaced with two different directional couplers that permit efficient, mode-specific coupling into and out of the resonator. The input coupler (coupler A) is designed to couple appreciably to both optical spatial modes, producing a characteristic cavity transmission (thru) spectrum with two distinct sets of resonant features (see in Fig. \ref{fig:deviceconcept}c). The broad (narrow) resonances correspond to the cavity modes produced by the antisymmetric (symmetric) optical spatial mode. Using the frequency selectivity of the resonator, the pump light ($\omega_{\rm p}$) is resonantly coupled into an antisymmetric cavity mode while the signal light ($\omega_{\rm s}$) is coupled into the symmetric cavity mode. Signal light circulating in the symmetric cavity mode exits the resonator through a mode-selective coupler (coupler B), which preferentially couples to the symmetric spatial mode (see Appendix B for details).

\begin{figure*}[ht]
\centering%\vspace{-10pt}
\includegraphics[width=\linewidth]{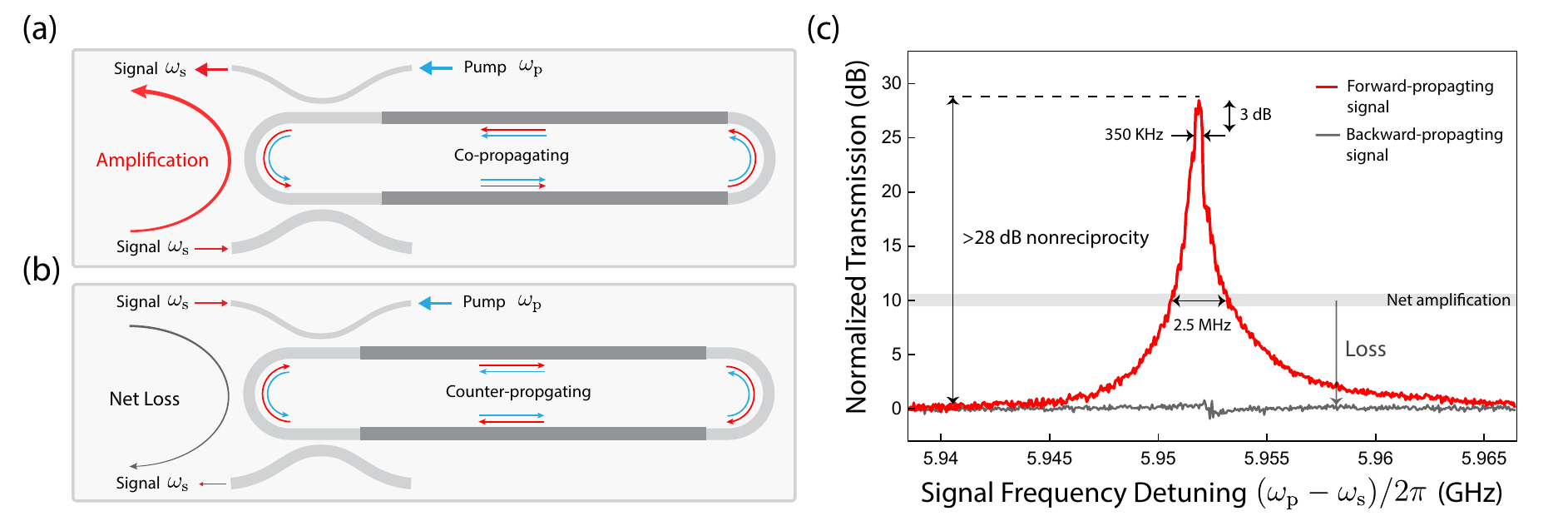}
\caption{Demonstration of unidirectional optical amplification and nonreciprocity (For experimental apparatus, see Appendix C1) (a) Experimental arrangement for directional amplification. Pump and signal waves are injected through respective multimode (top) and mode-specific (bottom) couplers such that they co-propagate (forward direction) within the resonator. This configuration allows pump and signal waves to nonlinearly couple through a stimulated forward inter-modal Brillouin process, yielding net amplification of the signal wave. (b) By contrast, a signal wave propagating in the opposite (backward) direction, does not experience Brillouin gain as a result of phase matching; the elastic wave that mediates forward inter-modal scattering is not phase-matched to the backward-scattering process. Thus, in this backward configuration, the signal wave experiences net loss resulting from linear transmission through the resonator.  (c) Experimental demonstration of unidirectional amplification. Signal transmission through the system in the forward (red; co-propagating with the pump) and backward (gray; counter-propagating with the pump) directions as a function of signal frequency detuing $\Omega/2 \pi$. This system yields a maximum 28 dB of nonreciprocity (with a FWHM of 350 KHz) and provides $>10$ dB of isolation over a 2.5 MHz bandwidth. }
\label{fig:nonreciprocal}
\end{figure*}

When the pump wave is resonant with an antisymmetric cavity mode ($\omega_{\rm p}=\omega_2^{m}$) and a symmetric cavity mode satisfies the Brillouin condition ($\omega_1^{n}=\omega_2^{m}-\Omega_{\rm B}$), signal light injected into the symmetric cavity mode ($\omega_1^{n}$) can experience resonantly enhanced Brillouin amplification.  Through experiments, we couple the pump and signal waves into the antisymmetric and symmetric cavity modes of the racetrack resonator via coupler A.  Within the resonator, the co-propagating pump and signal waves nonlinearly couple as they traverse the Brillouin-active regions of the racetrack, producing Brillouin energy transfer and single-sideband gain through stimulated inter-modal Brillouin scattering \cite{kittlaus2017chip}. As the pump power approaches the threshold for lasing, the degree of amplification is significantly enhanced (see Fig. \ref{fig:deviceconcept}d).

\subsection{Experimental results}

We characterize the resonantly enhanced Brillouin amplifier through nonlinear laser spectroscopy measurements using the setup diagrammed in Fig. \ref{fig:data}a.  All measurements are performed at room temperature and atmospheric pressure using 1.5 $\upmu$m wavelengths.  In this experimental scheme, light from a tunable laser is split along two paths; the upper path is used to synthesize the pump and signal waves, while the lower arm is used to create an optical local oscillator (LO) for heterodyne analysis of the emitted signal wave. The optical LO (lower arm) is generated by an acousto-optic modulator, which blueshifts the light by $\Delta=2 \pi \times 44$ MHz.  The upper path uses an intensity modulator (IM), erbium-doped fiber amplifier (EDFA), and variable optical attenuator (VOA) to synthesize pump and signal waves of a desired power and variable frequency detuning.  Pump and signal waves are then coupled on-chip through a grating coupler; the light is subsequently routed to the racetrack resonator through a single-mode waveguide. Signal light exiting the device is combined with the optical LO and measured using a high-speed photo-receiver for heterodyne spectral analysis. We sweep a microwave oscillator at $\Omega$ to synthesize the signal wave at $\omega_{\rm s}= \omega_{\rm p}-\Omega$, and synchronously detect at $\Omega+\Delta$ using a spectrum analyzer. By tracking at this offset frequency ($\Delta$), we are able to selectively detect the redshifted signal wave ($\omega_{\rm p}-\Omega$) without crosstalk from the blueshifted tone ($\omega_{\rm p}+\Omega$).

Using this experimental configuration, we sweep the laser wavelength, pump power, and signal-wave detuning to characterize the amplifier system.  In the limit of low pump power, no Brillouin gain is produced, and we recover the linear resonant response produced by the symmetric mode of the racetrack cavity.  When the pump power is increased, we observe a narrow gain peak at the Brillouin frequency atop the linear resonator response (see Fig. \ref{fig:deviceconcept}d).  As the pump power approaches the laser threshold power, we observe a dramatic increase in the degree of resonantly enhanced amplification, consistent with our theoretical predictions (see Fig. \ref{fig:data}c-d). Just below threshold, this resonantly enhanced interaction is sufficient to yield in excess of 30 dB of gain, representing more than 20 dB of net amplification after accounting for losses produced by linear transmission through the resonator (see Fig. \ref{fig:data}c).  The degree of amplification depends strongly on the mode-pair detuning (i.e., the frequency separation $\omega^{m}_2-\omega^{n}_1$) relative to the Brillouin frequency, as shown in Fig. \ref{fig:data}c.  In addition, we observe that the gain bandwidth scales inversely with the amplification as a result of gain narrowing, in agreement with theory (see Fig. \ref{fig:data}e).

Owing to the phase-matching characteristics of this inter-band Brillouin process, we show that this system yields unidirectional gain that results in a highly nonreciprocal response. This is because the phonon required for the resonantly enhanced Brillouin process mediates gain between co-propagating pump and signal waves, but does not produce Brillouin coupling between counter-propagating waves \cite{kang2011reconfigurable,poulton2012design}. We demonstrate these dynamics by coupling the pump wave into the anti-symmetric resonator mode with a counter-clockwise orientation; we then examine the reciprocity of the system by injecting the signal wave in the forward (Fig. \ref{fig:nonreciprocal}a) and backward (Fig. \ref{fig:nonreciprocal}b) directions such that the signal wave co- and counter-propagates with the pump wave within the resonator, respectively. When energy conservation is satisfied ($\omega_{\rm p}=\omega_{\rm s}+\Omega_{\rm B}$), the forward configuration yields net amplification (red) of the signal wave, while the backward configuration yields net loss (gray) as a result of linear transmission through the cavity in the absence of gain.  Using a fiber-coupled switch, we alternate between the forward and backward configurations while measuring the transmission as a function of signal-wave detuning. As shown in Fig. \ref{fig:nonreciprocal}c, these measurements reveal a peak optical nonreciprocity of 28 dB and a bandwidth of 2.5 MHz over which the system provides $>10$ dB of optical isolation with no insertion loss (see Appendix C1 for more details).

\subsection{Theory}

To understand our observations, we develop a mean-field analytical model that captures the salient amplification and noise dynamics of this resonantly enhanced amplification process (for detailed derivation, see Appendix A).  This model treats the pump, signal, and phonon fields as distinct modes that are nonlinearly coupled through stimulated inter-modal Brillouin scattering. Taking a Fourier transform in time allows us to solve for the output signal power spectrum $|S^{\rm out}[\omega_{\rm s}]|^2$ relative to the input signal power spectrum $|S^{\rm in}[\omega_{\rm s}]|^2$, yielding

\begin{equation}
\begin{aligned}
\frac{|S^{\rm out}[\omega_{\rm s}]|^2}{|S^{\rm in}[\omega_{\rm s}]|^2}=
 \left|\frac{\sqrt{\gamma_{\rm A,1}}\sqrt{\gamma_{\rm B,1}}}{-i(\omega_{\rm s}-\omega^n_1)+\frac{\gamma_{\rm tot,1}}{2}-\frac{G_{\rm B} P v_{\rm g,1}\Gamma/4}{i (\Omega_{\rm B}+\omega_{\rm s}-\omega_{\rm p})+\Gamma/2}}\right|^2,
\end{aligned}
\label{eq:gainm}
\end{equation}

\noindent where $\gamma_{\rm tot,1}$ is the total loss rate for the symmetric spatial mode (defined by $\gamma_{\rm tot,1}\equiv \gamma_{\rm A,1}+\gamma_{\rm B,1}+\alpha_1 v_{\rm g,1}$), $\gamma_{\rm (A,B),1}$ are the dissipation rates for the symmetric spatial mode due to couplers A and B ($\gamma_{\rm (A,B),1}\equiv -(2 v_{\rm g,1}/L)\ln{(1-\mu^2_{\rm (A,B),1})^{1/2}}$), $v_{\rm g,1}$ is the group velocity of the symmetric optical spatial mode, $\mu^2_{\rm (A,B),1}$ is the coupling constant of couplers A or B for the symmetric spatial mode, $L$ is the length of the racetrack resonator, $P$ is the intracavity pump power, $\alpha_1$ is the linear propagation loss of the symmetric spatial mode, $G_{\rm B}$ is the Brillouin gain coefficient, and $\Gamma$ is the dissipation rate for the acoustic field.

Equation \ref{eq:gainm} can be used to self-consistently predict the amplification and gain bandwidth of the system and is the basis for the theoretical trends plotted in Fig. \ref{fig:data}d-e. Using this framework, we also analyze the noise dynamics and gain depletion produced in this Brillouin amplifier (for detailed analysis of the noise figure and gain depletion, see Appendix A2-3). We note that Eq. \ref{eq:gainm} diverges at the laser threshold condition, which is an artifact of the stiff pump approximation in this analysis. While the equations are consistent with those describing parametric amplification in cavity-optomechanical systems \cite{safavi2011electromagnetically}, we note that this analysis requires a mean-field treatment of a distributed, heavily spatially damped phonon field, which is valid only in the presence of the pump and Stokes field (for further discussion, see Ref. \cite{otterstrom2018silicon}).

\section{Discussion}

In this paper, we have demonstrated that a resonant optical configuration can be used to dramatically enhance the stimulated inter-modal Brillouin scattering process, yielding record-high Brillouin gain and net amplification in an all-silicon chip-integrated system.  These results represent a 500-fold improvement in gain and more than a $60\times$ enhancement in net amplification relative to linear (non-resonant) devices of the same design \cite{kittlaus2017chip}. Building on this work, even greater degrees of optical amplification may be realized by increasing the passive signal transmission while maintaining a low laser threshold; this may be accomplished through further optimization of the mode-specific coupler design or multimode optomechanical waveguide.

As a byproduct of this resonantly enhanced interaction, we also show that this process yields characteristic narrowing of the gain bandwidth (from 10 MHz to sub-MHz) as the pump power approaches the laser threshold power.  While broadband amplification is desirable for many applications, the narrowband amplification produced through this resonantly enhanced system presents its own set of intriguing opportunities. In contrast to broadband amplification, in which spontaneous emission or scattering can produce substantial noise over an equally large bandwidth, the narrowband nature of the interaction yields low out-of-band noise (for details, see Appendix A2). In addition, the narrow, tailorable nature of the gain bandwidth could prove advantageous for many on-chip functionalities, including narrow-band optical and microwave-photonic filters \cite{tanemura2002narrowband,pant2014chip}, carrier recovery for microwave photonic signal processing \cite{giacoumidis2018chip}, and tunable time delay \cite{song2005observation,okawachi2005tunable,pant2012photonic,safavi2011electromagnetically}.

In addition, the unidirectional nature of this Brillouin amplifier opens the door to new types of all-silicon, chip-integrated nonreciprocal technologies. These nonreciprocal dynamics are closely related to those recently demonstrated in cavity-optomechanical and nonlinear optical systems, where time modulation produced through a parametric coupling can produce directional absorption or amplification \cite{sounas2017non,miri2017optical}. While similar behaviors have been demonstrated in glass micro-resonators \cite{kim2015non,dong2015brillouin,shen2016experimental,ruesink2016nonreciprocity,hua2016demonstration}, photonic crystal fibers \cite{kang2011reconfigurable}, and in silicon optomechanical crystals at cryogenic temperatures \cite{fang2017generalized}, this form of nonreciprocity has not previously been demonstrated in a silicon system at room temperature. Moreover, though narrowband in comparison, the level of nonreciprocity demonstrated here ($\sim 30$ dB) is competitive with that achievable using integrated magneto-optic \cite{shoji2008magneto,bi2011chip,huang2016electrically,pintus2019broadband} or acousto-optic strategies \cite{sohn2018time,kittlaus2018non}. While magneto-optic-based silicon-photonic isolator and circulator technologies are advancing steadily, they require complex fabrication techniques \cite{shoji2008magneto,bi2011chip,huang2016electrically,pintus2019broadband}. Acousto-optic strategies are also very promising; however, they have yet to achieve efficiencies necessary to produce optical isolation with low insertion losses \cite{sohn2018time,kittlaus2018non}.   As such, the optical nonreciprocity we demonstrate here---in an all-silicon device with no insertion loss---represents an important step towards practical isolator and circulator technologies in silicon photonics.

In summary, we have demonstrated record-high Brillouin gain and amplification in an integrated silicon photonic circuit.  This device is capable of delivering more than 30 dB of gain and 20 dB of net amplification, representing a 30-fold improvement over state-of-the-art performance \cite{kittlaus2016large}.  Moreover, we show that this phase-matched stimulated Brillouin process is intrinsically unidirectional, yielding more than 28 dB of nonreciprocal contrast between forward- and backward-propagating waves. These results represent a important milestone for Brillouin-based amplifier and isolator technologies in silicon photonics and open the door to new schemes for high-performance microwave photonic filtering, tunable time delay, and injection-locked Brillouin lasers.

\section*{Funding Information}

This material is based upon work supported by the Packard Fellowship for Science and Engineering, the National Science Foundation Graduate Research Fellowship under Grant No. DGE1122492 (N.T.O.), and the Laboratory Directed Research and Development program at Sandia National Laboratories. Sandia National Laboratories is a multi-program laboratory managed and operated by National Technology and Engineering Solutions of Sandia, LLC., a wholly owned subsidiary of Honeywell International, Inc., for the U.S. Department of Energy's National Nuclear Security Administration under contract DE-NA-0003525. This paper describes objective technical results and analysis. Any subjective views or opinions that might be expressed in the paper do not necessarily represent the views of the U.S. Department of Energy, the National Science Foundation, or the United States Government.

\section*{Acknowledgements}

We thank Prashanta Kharel for valuable discussions and for assistance in developing the experimental apparatus.

\newpage
\newpage

\onecolumngrid

\newpage
\appendix
\maketitle
\tableofcontents
\newpage

\section{Dynamics of resonantly enhanced Brillouin amplifier}
\subsection{Mean-field analysis}

In this section, we describe a simple mean-field treatment of the amplifier dynamics that is sufficient to capture the salient behavior of the resonantly enhanced inter-modal Brillouin process.  Combing mean-field analysis of the full spatio-temporal dynamics of a resonant Brillouin system (see Supplementary Materials of Ref. \cite{otterstrom2018silicon}) with standard coupled-mode theory \cite{little1997microring}, we can express the equations of motion as

\begin{equation}
\begin{aligned}
\dot{a}_{\rm s}(t)&=-i \omega^{n}_{1} a_{\rm s}(t) - \frac{\gamma_{\rm tot,1}}{2}a_{\rm s}(t) -i g^{*} a_{\rm p}(t)b^{\dagger}(t)+\sqrt{\gamma_{\rm A,1}}S^{\rm in}(t)+\sqrt{\gamma_{\rm A,1}}S_{\rm A}^{\rm N}(t)+\sqrt{\gamma_{\rm B,1}}S_{\rm B}^{\rm N}(t)+\sqrt{\gamma_{\rm R,1}}S_{\rm R}^{\rm N}(t)\\
\dot{b}(t)&=-i \Omega_{\rm B} b(t) - \frac{\Gamma}{2} b(t)-ig^{*} a_{\rm p}(t) a^{\dagger}_{\rm s}(t)+\eta(t)\\
\dot{a}_{\rm p}(t)&=-i \omega^{m}_2 a_{\rm p}(t)-\frac{\gamma_{\rm tot,2}}{2}a_{\rm p}(t)-i g a_{\rm s}(t)b(t)+\sqrt{\gamma_{\rm A,2}}S^{\rm in}_{\rm p}(t)+\sqrt{\gamma_{\rm tot,2}}S_{\rm tot,2}^{\rm N}(t).
\end{aligned}
\label{eq:eqmot}
\end{equation}

Here, $a_{\rm s}(t)$, $b(t)$, and $a_{\rm p}(t)$ are the coupled-mode amplitudes for the signal, phonon, and pump fields, respectively, with units of $[\sqrt{\textup{number}}]$; $S^{\rm in}(t)$ and $S^{\rm in}_{\rm p}(t)$ represent the input signal-wave and pump-wave fields (with units $[\sqrt{\textup{number}\times \textup{Hz}}]$); g is the acousto-optic coupling; $\eta(t)$ is the mechanical stochastic driving term that is consistent with the fluctuation-dissipation theorem, with a two-time correlation function of $\langle \eta^{\dagger}(t^\prime) \eta(t) \rangle = \Gamma (n_{\rm th}+1/2) \delta(t-t^\prime)$, where $n_{\rm th}$ is the thermal occupation of the phonon field; $S^{\rm N}_{\rm(A,B,R)}(t)$ represent the vacuum fluctuations associated with the optical losses produced by (A) coupler A, (B) coupler B, and (R) the intrinsic loss of the of the racetrack resonator, each with a two-time correlation function given by $\langle S^{\rm N \dagger}_{\rm(A,B,R)}(t^\prime) S^{\rm N}_{\rm(A,B,R)}(t)\rangle=(1/2)\delta(t-t^\prime)$; $S^{\rm N}_{\rm tot,2}(t)$ represents the vacuum fluctuations associated with the total loss experienced by an antisymmetric optical spatial mode; $\gamma_{\rm tot,(1,2)}$  is the total loss rate for the symmetric and antisymmetric spatial modes (defined by $\gamma_{\rm tot,(1,2)}\equiv \gamma_{\rm A,(1,2)}+\gamma_{\rm B,(1,2)}+\alpha_{(1,2)} v_{\rm g,(1,2)}$); $\gamma_{\rm (A,B),(1,2)}$ are the dissipation rates for the symmetric and antisymmetric spatial modes due to couplers A and B ($\gamma_{(A,B),(1,2)}\equiv -(2 v_{\rm g,(1,2)}/L)\ln{(1-\mu^2_{\rm (A,B),(1,2)})^{1/2}}$) while $\gamma_{\rm R,1}$ represents the internal losses of the racetrack resonator for the symmetric mode ($\gamma_{\rm R,1}=\alpha_{1} v_{\rm g,1}$); $v_{\rm g,(1,2)}$ is the group velocity of the symmetric and antisymmetric optical spatial modes; $L$ is the length of the racetrack resonator; $\alpha_{(1,2)}$ is the linear propagation loss of the symmetric and antisymmetric spatial mode; $\Gamma$ is the dissipation rate for the acoustic field.

Using the rotating-wave approximation, we center each field around DC, yielding

\begin{equation}
\begin{aligned}
\dot{\bar{a}}_{\rm s}(t)&=i(\omega_{\rm s}- \omega^{n}_{1}) \bar{a}_{\rm s}(t) - \frac{\gamma_{\rm tot,1}}{2}\bar{a}_{\rm s}(t) -i g^{*} \bar{a}_{\rm p}(t)\bar{b}^{\dagger}(t)+\sqrt{\gamma_{\rm A,1}}\bar{S}^{\rm in}(t)+\sqrt{\gamma_{\rm A,1}}\bar{S}_{\rm A}^{\rm N}(t)+\sqrt{\gamma_{\rm B,1}}\bar{S}_{\rm B}^{\rm N}(t)+\sqrt{\gamma_{\rm R,1}}\bar{S}_{\rm R}^{\rm N}(t)\\
\dot{\bar{b}}(t)&=i (\Omega-\Omega_{\rm B}) \bar{b}(t) - \frac{\Gamma}{2} \bar{b}(t)-ig^{*} \bar{a}_{\rm p}(t) \bar{a}^{\dagger}_{\rm s}(t)+\bar{\eta}(t)\\
\dot{\bar{a}}_{\rm p}(t)&=i(\omega_{\rm p}- \omega^{m}_2) \bar{a}_{\rm p}(t)-\frac{\gamma_{\rm tot,2}}{2}\bar{a}_{\rm p}(t)-i g \bar{a}_{\rm s}(t)\bar{b}(t)+\sqrt{\gamma_{\rm A,2}}\bar{S}^{\rm in}_{\rm p}(t)+\sqrt{\gamma_{\rm tot,2}}\bar{S}_{\rm tot,2}^{\rm N}(t),
\end{aligned}
\label{eq:rwa}
\end{equation}

\noindent
where $a_{\rm s}(t)=\bar{a}_{\rm s}(t)\exp{(-i \omega_{\rm s} t)}$, $a_{\rm p}(t)=\bar{a}_{\rm p}(t)\exp{(-i \omega_{\rm p} t)}$, and $b(t)=\bar{b}(t)\exp{(-i \Omega t)}$); $\bar{\eta}(t)$, $\bar{S}^{\rm in}(t)$,$\bar{S}^{\rm N}_{(A,B,R)}(t)$, $\bar{S}_{\rm tot,2}^{\rm N}(t)$, and $\bar{S}^{\rm in}_{\rm p}(t)$ are shifted in like manner.  We also note that the above requires $\omega_{\rm p}=\omega_{\rm s}+\Omega$, consistent with our experimental scheme.

For simplicity, we treat the pump wave as undepleted (i.e., $\dot{\bar{a}}_{\rm p}=0$ and $\bar{a}_{\rm p}(t)=\bar{a}_{\rm p}$) and ignore the vacuum fluctuations associated with the pump. We next take a Fourier transform of Eq. \ref{eq:rwa} in time, yielding

\begin{equation}
\begin{aligned}
-i \omega \bar{a}_{\rm s}[\omega]&=i(\omega_{\rm s}- \omega^{n}_{1}) \bar{a}_{\rm s}[\omega] - \frac{\gamma_{\rm tot,1}}{2}\bar{a}_{\rm s}[\omega] -i g^{*} \bar{a}_{\rm p}\bar{b}^{\dagger}[\omega]+\sqrt{\gamma_{\rm A,1}}\bar{S}^{\rm in}[\omega]+\sqrt{\gamma_{\rm A,1}}\bar{S}_{\rm A}^{\rm N}[\omega]+\sqrt{\gamma_{\rm B,1}}\bar{S}_{\rm B}^{\rm N}[\omega]+\sqrt{\gamma_{\rm R,1}}\bar{S}_{\rm R}^{\rm N}[\omega]\\
-i \omega \bar{b}[\omega]&=i (\Omega-\Omega_{\rm B}) \bar{b}[\omega] - \frac{\Gamma}{2} \bar{b}[\omega]-i g^{*} \bar{a}_{\rm p} \bar{a}^{\dagger}_{\rm s}[\omega]+\bar{\eta}[\omega].\\
\end{aligned}
\label{eq:pd}
\end{equation}

\noindent Next, we solve these two coupled equations, yielding

\begin{equation}
\begin{aligned}
\bar{a}_{\rm s}[\omega]&=\frac{1}{-i(\omega+\omega_{\rm s}-\omega^n_1)+\frac{\gamma_{\rm tot,1}}{2}-\frac{|g|^2 |\bar{a}_{\rm p}|^2}{-i(\omega+\Omega-\Omega_{\rm B})+\frac{\Gamma}{2}}}\Bigg[\sqrt{\gamma_{\rm A,1}}\bar{S}^{\rm in}[\omega]\\&+\frac{-i g^{*}\bar{a}_{\rm p}\bar{\eta}[\omega]}{-i(\omega+\Omega-\Omega_{\rm B})+\frac{\Gamma}{2}}+\sqrt{\gamma_{\rm A,1}}\bar{S}_{\rm A}^{\rm N}[\omega]+\sqrt{\gamma_{\rm B,1}}\bar{S}_{\rm B}^{\rm N}[\omega]+\sqrt{\gamma_{\rm R,1}}\bar{S}_{\rm R}^{\rm N}[\omega] \Bigg].
\end{aligned}
\label{eq:sol1}
\end{equation}
\noindent

We proceed by transforming out of the rotating frame and calculating the output signal spectrum by noting that $S^{\rm out}[\omega_{\rm s},\omega]=\sqrt{\gamma_{\rm B,1}} a_{\rm s}[\omega_{\rm s},\omega]-S_{\rm B}^{\rm N}[\omega]$, yielding

\begin{equation}
\begin{aligned}
S^{\rm out}[\omega_{\rm s},\omega]&=T[\omega]\Bigg[\sqrt{\gamma_{\rm A,1}}S^{\rm in}[\omega_{\rm s}+\omega]+\frac{-i g^{*}\bar{a}_{\rm p}\bar{\eta}[\omega]}{-i(\omega+\Omega-\Omega_{\rm B})+\frac{\Gamma}{2}} +\sqrt{\gamma_{\rm A,1}}\bar{S}_{\rm A}^{\rm N}[\omega]+\sqrt{\gamma_{\rm B,1}}\bar{S}_{\rm B}^{\rm N}[\omega]+\sqrt{\gamma_{\rm R,1}}\bar{S}_{\rm R}^{\rm N}[\omega]\Bigg]-\bar{S}_{\rm B}^{\rm N}[\omega],
\end{aligned}
\label{eq:sol3}
\end{equation}

\noindent
where we define the resonator transfer function ($T[\omega]$) as

\begin{equation}
\begin{aligned}
T[\omega_{\rm s},\omega]\equiv\frac{\sqrt{\gamma_{\rm B,1}}}{-i(\omega+\omega_{\rm s}-\omega^n_1)+\frac{\gamma_{\rm tot,1}}{2}-\frac{|g|^2 |\bar{a}_{\rm p}|^2}{-i(\omega+\omega_{\rm p}-\omega_{\rm s}-\Omega_{\rm B})+\frac{\Gamma}{2}}}.
\end{aligned}
\label{eq:sol2}
\end{equation}

We proceed by computing the power spectral density of the output signal.  This yields

\begin{equation}
\begin{aligned}
|S^{\rm out}[\omega_{\rm s},\omega]|^2&=
 |T[\omega_{\rm s},\omega]|^2\Bigg[\underbrace{\gamma_{\rm A, 1}|S^{\rm in}[\omega_{\rm s}+\omega]|^2}_{{\rm Input\:Signal}}+  \underbrace{\frac{G_{\rm B} P v_{\rm g, 1} \Gamma^2 (n_{\rm th}+1/2)}{4((\omega+\omega_{\rm p}-\omega_{\rm s}-\Omega_{\rm B})^2+(\frac{\Gamma}{2})^2)}}_{{\rm Thermal-mechanical\:Noise}} \\&+ \underbrace{\gamma_{\rm A,1}|\bar{S}_{\rm A}^{\rm N}[\omega]|^2+|\sqrt{\gamma_{\rm B,1}}-\frac{1}{T[\omega_{\rm s},\omega]}|^2|\bar{S}_{\rm B}^{\rm N}[\omega]|^2+\gamma_{\rm R,1}|\bar{S}_{\rm R}^{\rm N}[\omega]|^2}_{{\rm Optical\:Vacuum\:Fluctuations}}\Bigg],
\end{aligned}
\label{eq:psd}
\end{equation}

where the Brillouin gain coefficient $G_{\rm B}$ is defined as

\begin{equation}
\begin{aligned}
G_{\rm B} = \frac{4 |g|^2 |a_{\rm p}|^2}{P \Gamma v_{\rm g,1}}
\end{aligned}
\label{eq:def}
\end{equation}

Equation \ref{eq:psd} reveals three distinct terms in the output power spectrum; the first term gives the resonant amplification of the input signal, the second arises from thermal-mechanical noise, and the last results from optical vacuum fluctuations. We note here that the cross-terms vanish because the signal and noise sources are uncorrelated (e.g., $\eta^{\dagger}[\omega]S^{\rm in}[\omega_{\rm s}+\omega]=FT[\langle \eta^{\dagger}(t+\tau)S^{\rm in}(t)\rangle]=0$).

We first calculate the signal-wave amplification ($|S^{\rm out}[\omega_{\rm s}]|^2/|S^{\rm in}[\omega_{\rm s}]|^2$) by analyzing the spectrum at the center frequency of the input signal (i.e., $\omega=0$) and ignoring the noise terms, which yields

\begin{equation}
\begin{aligned}
\frac{|S^{\rm out}[\omega_{\rm s}]|^2}{|S^{\rm in}[\omega_{\rm s}]|^2}=
 \left|\frac{\sqrt{\gamma_{A,1}}\sqrt{\gamma_{B,1}}}{-i(\omega_{\rm s}-\omega^n_1)+\frac{\gamma_{\rm tot,1}}{2}-\frac{G_{\rm B} P v_{\rm g,1}\Gamma/4}{i (\Omega_{\rm B}+\omega_{\rm s}-\omega_{\rm p})+\Gamma/2}}\right|^2.
\end{aligned}
\label{eq:gain}
\end{equation}

\noindent This is the central result of this section. We consider the effects of noise in the following section.

\begin{figure*}[!b]
\centering%\vspace{-10pt}
\includegraphics[width=\linewidth]{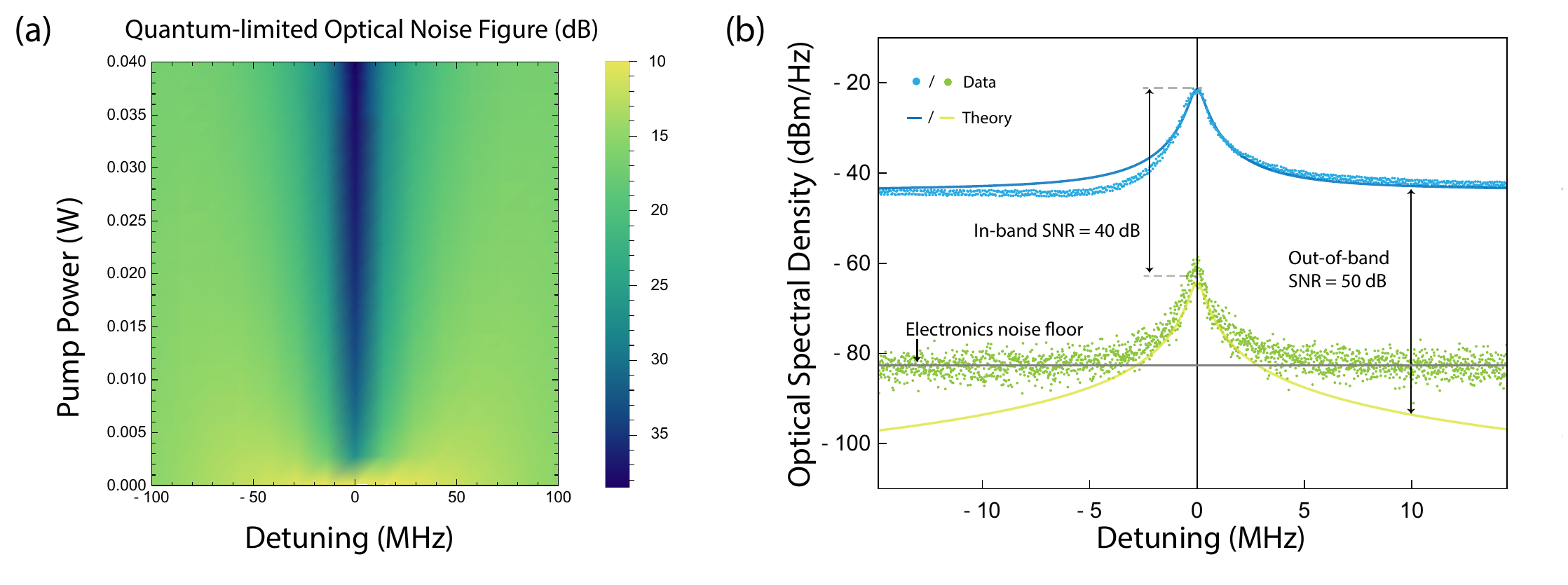}
\caption{(a) Theoretical quantum-limited optical noise figure (relative to quantum-limited input signal) as a function of detuning and pump power, obtained from Eq. \ref{eq:NFF}. (b)  Output signal (blue), Brillouin noise (light green), and respective theory curves (obtained from Eq. \ref{eq:psd}, with $P_p=0.93P_{\rm th}$, where $P_{\rm th}$ is the laser threshold power). Data recorded with $\rm RBW = 50$ KHz. Theory curves obtained from Eq. \ref{eq:psd}.  }
\label{fig:noise}
\end{figure*}

\subsection{Noise dynamics}

Building on this framework, we can also analyze the quantum-limited noise figure for the amplifier system.  The noise figure is defined by 

\begin{equation}
\begin{aligned}
\textup{NF}=\frac{\textup{SNR}_1}{\textup{SNR}_2},
\end{aligned}
\label{eq:nf}
\end{equation}

\noindent
where $\textup{SNR}_1$ is the input signal-to-noise ratio (SNR) and $\textup{SNR}_2$ is the output signal-to-noise ratio.

As an upper bound on the noise figure, we assume a quantum-limited input signal (i.e., signal relative to the zero-point background; for more details see Ref. \cite{haus1998noise,bristiel2006new}).  We consider the output SNR over a vanishingly small bandwidth $2 \Delta \omega$  ($\Delta \omega\ll \Gamma$) centered around $\omega=0$. The noise sources are given by the amplified thermal-mechanical and optical vacuum fluctuations (see Eq. \ref{eq:psd}), while the output signal is given by the amplified input signal. For simplicity, we assume the following functional form for the input signal spectrum: $|S^{\rm in}[\omega]|^2=|S^{\rm in}|^2 \delta(\omega-\omega_{\rm s})$, such that 

\begin{equation}
\begin{aligned}
\textup{SNR}_2&=\frac{\int_{-\Delta \omega}^{\Delta \omega}d\omega \gamma_{\rm A,1} |S^{\rm in}|^2 \delta(\omega-\omega_{\rm s})}{\int_{-\Delta \omega}^{\Delta \omega}d\omega N_{\rm b}[\omega_{\rm s},\omega]+\gamma_{\rm A,1}|\bar{S}_{\rm A}^{\rm N}[\omega]|^2+|\sqrt{\gamma_{\rm B,1}}-\frac{1}{T[\omega_{\rm s},\omega]}|^2|\bar{S}_{\rm B}^{\rm N}[\omega]|^2+\gamma_{\rm R,1}|\bar{S}_{\rm R}^{\rm N}[\omega]|^2}\\
&\simeq \frac{\gamma_{\rm A,1}|S^{\rm in}|^2}{2 \Delta \omega \Big[ N_{\rm b}[\omega_{\rm s},0]+\frac{1}{2}\Big( \gamma_{\rm A,1}+|\sqrt{\gamma_{\rm B,1}}-\frac{1}{T[\omega_{\rm s},0]}|^2+\gamma_{\rm R,1} \Big) \Big]}
\end{aligned}
\label{eq:snrOut}
\end{equation}

\noindent
where $N_{\rm b}[\omega_{\rm s},\omega]$ is defined by 

\begin{equation}
\begin{aligned}
N_{\rm b}[\omega_{\rm s},\omega]\equiv\frac{G_{\rm B} P v_{\rm g, 1} \Gamma^2 (n_{\rm th}+1/2)}{4((\omega+\omega_{\rm p}-\omega_{\rm s}-\Omega_{\rm B})^2+(\frac{\Gamma}{2})^2)}.
\end{aligned}
\label{eq:Nb}
\end{equation}

\noindent
Here, we have used the correlation properties of the optical vacuum fluctuations to compute each power spectral density (i.e., $|\bar{S}_{\rm A}^{\rm N}[\omega]|^2=\int^{\infty}_{-\infty}d\tau \langle S^{\rm N \dagger}_{\rm(A,B,R)}(t+\tau) S^{\rm N}_{\rm(A,B,R)}(t)\rangle e^{i\omega \tau}=\int^{\infty}_{-\infty}d\tau(1/2)\delta(\tau)e^{i\omega \tau}=1/2)$ ).

We next calculate the input SNR. To calculate an upper bound on the noise figure, we assume a quantum-limited input signal (i.e., limited only by optical vacuum fluctuations).  In this case, the input SNR is given by

\begin{equation}
\begin{aligned}
\textup{SNR}_1&=\frac{\int_{-\Delta \omega}^{\Delta \omega}d\omega |S^{\rm in}|^2 \delta(\omega-\omega_{\rm s})}{\int_{-\Delta \omega}^{\Delta \omega}d\omega |\bar{S}_{\rm A}^{\rm N}[\omega]|^2}\\
&\simeq \frac{|S^{\rm in}|^2}{\Delta \omega }.
\end{aligned}
\label{eq:snrIn}
\end{equation}

Now, combining Eq. \ref{eq:snrOut} with Eq. \ref{eq:snrIn}, we calculate the noise figure of the system, yielding

\begin{equation}
\begin{aligned}
\textup{NF}&= \frac{ \Big[ 2N_{\rm b}[\omega_{\rm s},0]+\Big( \gamma_{\rm A,1}+|\sqrt{\gamma_{\rm B,1}}-\frac{1}{T[\omega_{\rm s},0]}|^2+\gamma_{\rm R,1} \Big) \Big]}{\gamma_{\rm A,1}}\\
&=\frac{\Big[ \frac{G_{\rm B} P v_{\rm g, 1} \Gamma^2 (n_{\rm th}+1/2)}{2((\omega_{\rm p}-\omega_{\rm s}-\Omega_{\rm B})^2+(\frac{\Gamma}{2})^2)}+\Big( \gamma_{\rm A,1}+|\sqrt{\gamma_{\rm B,1}}-\frac{-i(\omega_{\rm s}-\omega^n_1)+\frac{\gamma_{\rm tot,1}}{2}-\frac{|g|^2 |a_{\rm p}|^2}{-i(\omega_{\rm p}-\omega_{\rm s}-\Omega_{\rm B})+\frac{\Gamma}{2}}}{\sqrt{\gamma_{\rm B,1}}}|^2+\gamma_{\rm R,1} \Big) \Big] }{\gamma_{\rm A,1}}
\end{aligned}
\label{eq:NFF}
\end{equation}

We note that the noise figure is signal independent, as it should be \cite{haus1998noise}.  Figure \ref{fig:noise}a plots the noise figure (Eq. \ref{eq:NFF}) as a function of pump power and detuning from resonance. In the limit of low pump powers or large detunings, Eq. \ref{eq:NFF} converges to a value that is consistent with the loss produced by linear transmission through the system. When the pump power is increased, additional noise is imparted through spontaneous Brillouin scattering. As the pump power approaches the laser threshold, the quantum-limited noise figure of this silicon Brillouin amplifier system is approximately 38.5 dB relative to the zero-point background. This value is on par with the noise figure of conventional Brillouin amplifiers based on linear waveguides (which can range anywhere between 25 dB to 45 dB depending on the Brillouin frequency). In Brillouin amplifiers (such as the present system), the spontaneous Brillouin emission is narrowly centered around the Brillouin frequency (within $\sim 10$ MHz), and as such, this system exhibits low out-of-band noise; analyzing the performance of this system over larger bandwidths ($\sim 300 $ MHz), for example, yields an effective noise figure that is significantly reduced ($\textup{NF}\approx 15$ dB).  Moreover, in many practical applications, the input signal typically possesses additional noise---such as relative intensity noise (RIN), amplified spontaneous emission (ASE), or relaxation oscillations---whose magnitude far exceeds the zero-point background \cite{urick2015fundamentals}. Thus, while the quantum-limited noise figure we derive here provides a well-defined upper bound (worst case), the effective noise figure will be more favorable for a large range of practical applications.

\subsubsection{Experimental measure of noise}

We investigate the noise properties of this resonantly enhanced Brillouin amplifier through optical heterodyne laser spectroscopy using the apparatus diagrammed in Fig. 2a of the main text.  Figure \ref{fig:noise}b plots the transmitted signal wave (blue) and the spontaneous optical spectrum (green) versus frequency detuning (relative to the Brillouin frequency $\Omega_{\rm B}$), demonstrating good agreement with theoretical predictions (see Eq. \ref{eq:psd}). These results highlight the low out-of-band noise properties of the Brillouin amplifier system.

\begin{figure*}[!b]
\centering%\vspace{-10pt}
\includegraphics[width=.9\linewidth]{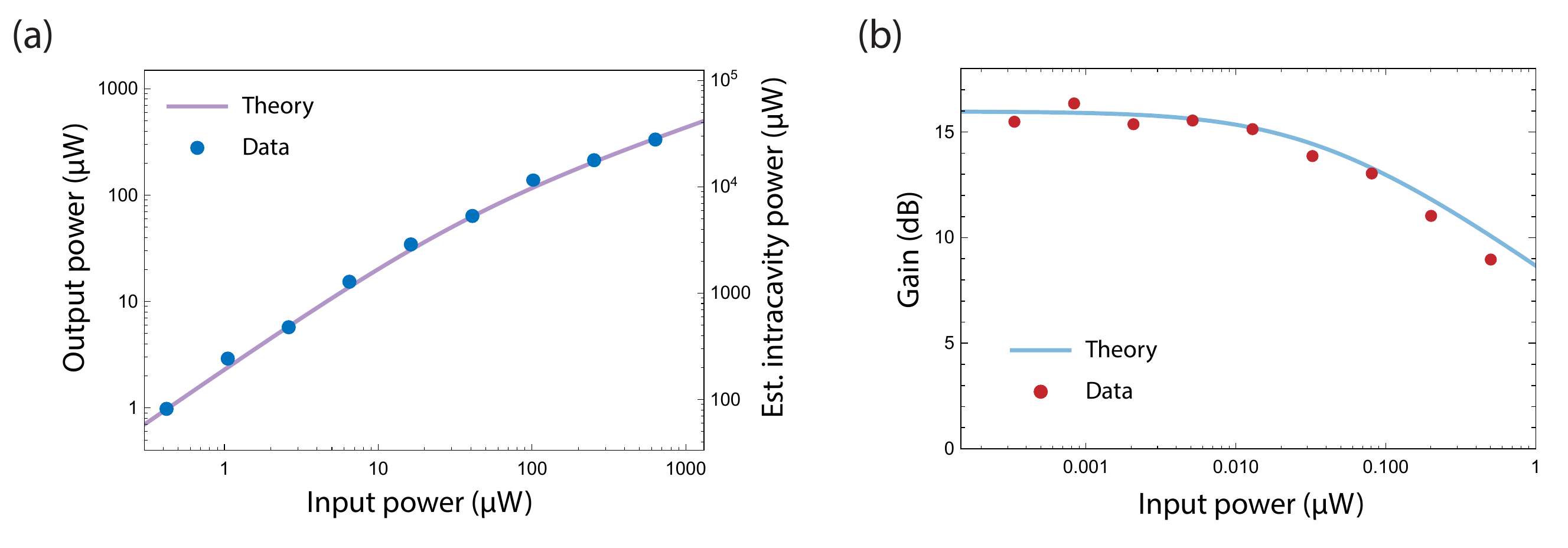}
\caption{(a) Output signal wave power as a function of input signal wave power. (b) Gain versus input power.  Theoretical curves are obtained by numerically solving Eq. \ref{eq:transend} (with $P=0.86P_{\rm th}$, $\mu^2_{\rm A,1}=0.16$, and $\mu^2_{\rm B,1}=0.012$) }
\label{fig:depletion}
\end{figure*}

\subsection{Gain depletion}

In this section, we explore the output signal saturation that arises from pump depletion within the resonator. We begin with the mean-field equations of motion under the rotating-wave approximation (Eq. \ref{eq:rwa}).  Here, we no longer assume a stiff pump; instead, we adiabatically eliminate the pump dynamics, in accodance with the separation of time scales ($\gamma_{\rm p}\gg \gamma_{\rm s} \gg \Gamma$).  

In this case, the slowly varying amplitude of the pump wave is given by

\begin{equation}
\begin{aligned}
\bar{a}_{\rm p}(t)=\frac{2}{\gamma_{\rm tot,2}}\big[-i g \bar{a}_{\rm s}(t) \bar{b}(t) + \sqrt{\gamma_{\rm A,2}} \bar{S}^{\rm in}_{\rm p}(t) \big].
\end{aligned}
\label{eq:adp}
\end{equation}

Next, we substitute Eq. \ref{eq:adp} into Eq. \ref{eq:rwa}, ignore thermal and quantum fluctuations, and solve for the steady-state behavior of the system, yielding the following transcendental equation for the signal wave field:

\begin{equation}
\begin{aligned}
\bar{a}_{\rm s}=\frac{\sqrt{\gamma_{\rm A,1}}\bar{S}^{\rm in}_{\rm s}}{\Bigg[ -i (\omega_{\rm s}-\omega^{n}_{\rm 1})+\frac{\gamma_{\rm tot,1}}{2}+\frac{G_{\rm B}^2 \hbar \omega_{\rm p} P v^2_{\rm g,1} v_{\rm g,2} \Gamma^2/(8 \gamma_{\rm tot,2} L)}{(\omega_{\rm p}-\omega_{\rm s}-\Omega_{\rm B})^2+(\Gamma/2+G_{\rm B} \hbar \omega_{\rm p} v_{\rm g,1} v_{\rm g,2} \Gamma |\bar{a}_{\rm s}|^2/(2 \gamma_{\rm tot,2} L))^2}-\frac{G_{\rm B} P v_{\rm g,1} \Gamma/4}{i (\omega_{\rm p}-\omega_{\rm s}- \Omega_{\rm B})+\Gamma/2+G_{\rm B} \hbar \omega_{\rm p} v_{\rm g,1} v_{\rm g,2} \Gamma |\bar{a}_{\rm s}|^2/(2 \gamma_{\rm tot,2} L)} \Bigg]}.
\end{aligned}
\label{eq:transend}
\end{equation}

\noindent
As this equation becomes analytically intractable, we proceed by solving Eq. \ref{eq:transend} numerically.

We also explore this gain depletion behavior through a set of systematic experimental studies.  This approach involves analyzing the transmitted signal-wave spectrum versus frequency over a range of input signal-wave powers; data from these measurements are presented in Fig. \ref{fig:depletion}.  Through these measurements, we observe a clear reduction in the gain as we increase the input signal wave power, in good agreement with our pump-depletion-based theoretical predictions.

\section{Passive resonator properties}

\begin{figure*}[ht]
\centering%\vspace{-10pt}
\includegraphics[width=.95\linewidth]{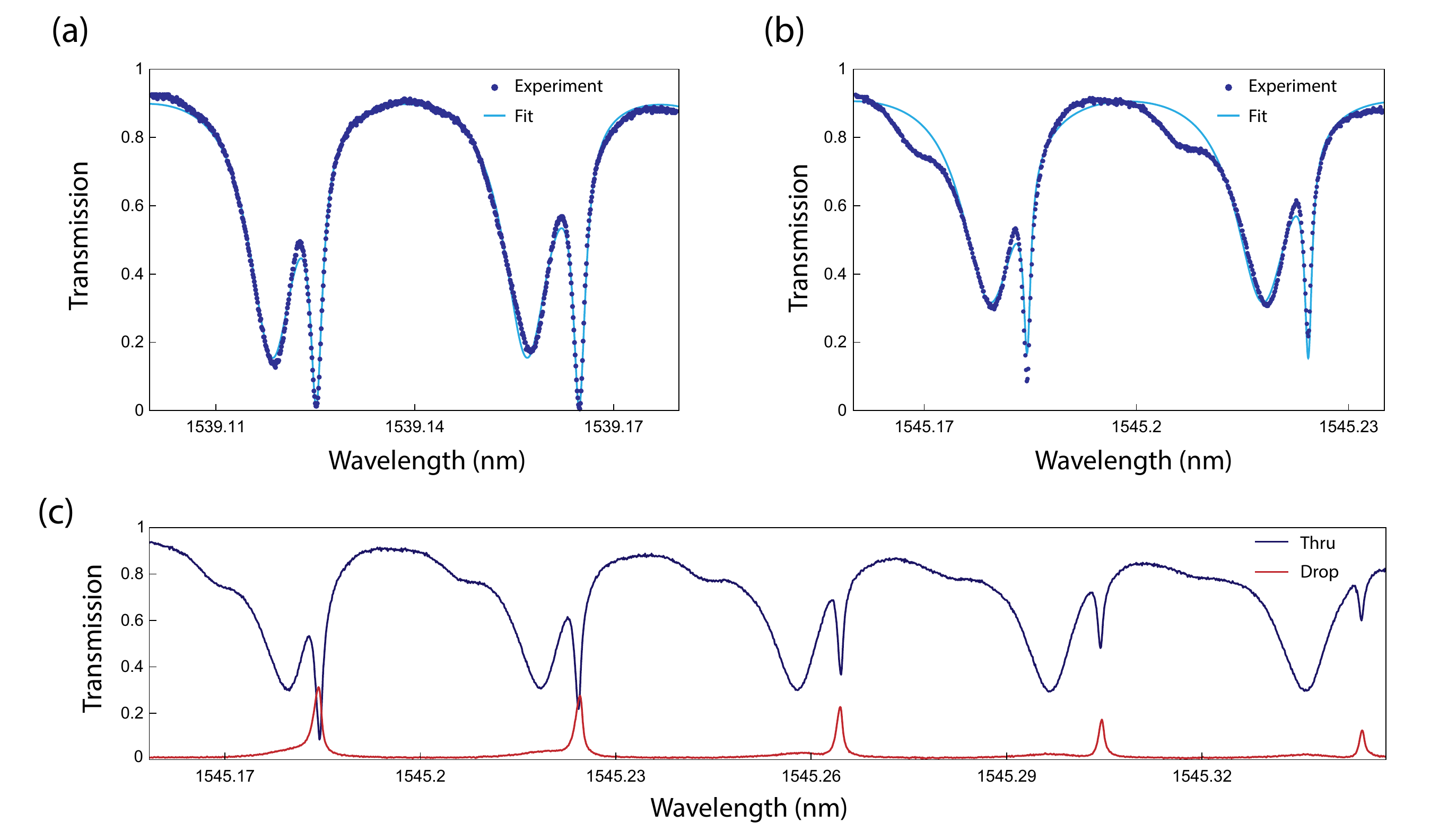}
\caption{(a)-(b) Linear (thru) transmission spectra centered around two distinct mode pairs that satisfy the Brillouin condition ($\omega_1^{n}=\omega_2^{m}-\Omega_{\rm B}$).  Narrow (broad) resonances correspond to the symmetric (antisymmetric) spatial modes of the multimode waveguide. Fits are obtained using the multimode ring resonantor theory presented in the Supplementary Materials of Ref. \cite{otterstrom2018silicon}.  (c) Transmission spectra measured at the thru and drop ports over a larger wavelength range. }
\label{fig:ring}
\end{figure*}

In this section, we present a brief explanation of the passive resonator properties. The multi-spatial mode racetrack cavity produces two distinct sets of cavity modes, yielding a characteristic multimode transmission spectrum (see Fig. \ref{fig:ring}c).  We measure the linear transmission response of the cavity using a swept 1.55 $\upmu$m laser (Agilent 81600B; sweep rate set to 5 nm/s) at low powers ($\sim120$ $\upmu$W on-chip). Fig. \ref{fig:ring}a-b plots the measured transmission (thru) spectra  centered around two distinct mode pairs that satisfy the Brillouin condition ($\omega_1^{n}=\omega_2^{m}-\Omega_{\rm B}$), while Fig. \ref{fig:ring}c shows the transmission spectra measured at the thru and drop ports over a larger wavelength range. Fitting these transmission spectra to a multimode resonator model (see Supplmentary Materials of Ref. \cite{otterstrom2018silicon}) allows us to obtain important parameters of the system, including the multimode coupling coefficients, group velocities, and linear propagation losses (see Table \ref{tab:par}).  This information is then used to model the amplifier dynamics (for example, see theory curve in Fig. 2d-e of the main text).

\section{Experimental considerations and data analysis}
\subsection{Nonreciprocal amplification characterization}

\begin{figure*}[ht]
\centering%\vspace{-10pt}
\includegraphics[width=.75\linewidth]{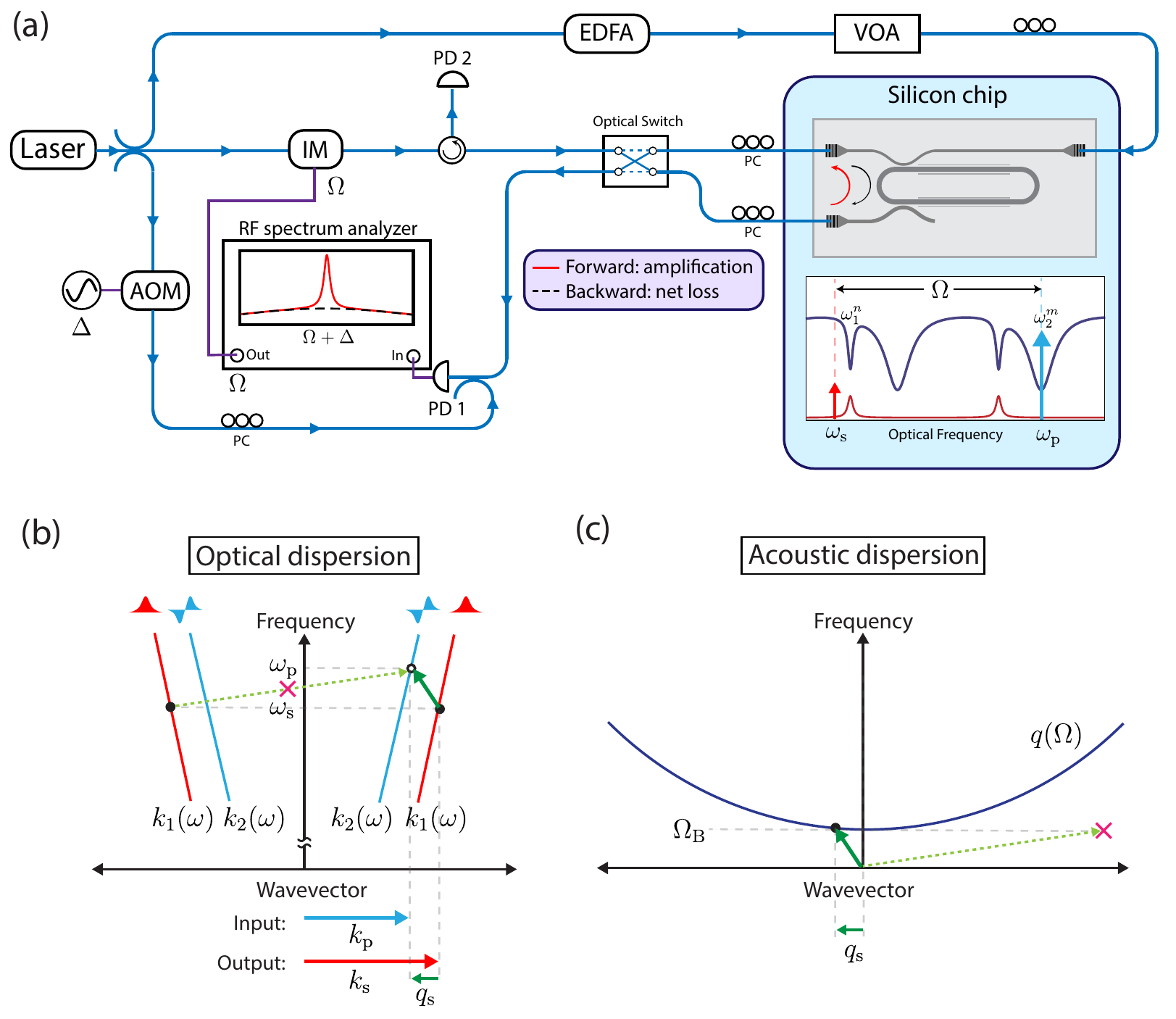}
\caption{Nonreciprocal optical amplification measurements. (a) Diagram of the experimental apparatus used to characterize the nonreciprocal nature of the resonantly enhanced Brillouin amplifier. Laser light is split along three paths. The first (top) is used to generate a pump wave of a desired optical power, which is subsequently coupled on-chip. The second arm synthesizes the signal wave with the desired frequency detuning ($\Omega=\omega_{\rm p}-\omega_{\rm s}$) using an intensity modulator (IM). Finally, the third creates an optical local oscillator (LO) using an acousto-optic modulator (AOM), which blueshifts the light by $\Delta=44$ MHz.  An optical switch directs the signal wave light on-chip such that it either co- or counter-propagates with the pump wave within the resonator.  After passing through the device, the signal wave is coupled through the drop port and off-chip, where it is combined with the blueshifted LO and measured on a high-speed photodector (PD 1). (b)-(c) diagrams illustrating energy and phase-matching conditions imposed by the optical and acoustic dispersion relations. In this diagram, the dark green arrow represents the frequency and wavevector of the phonon that mediates stimulated inter-modal Brillouin scattering, while the dotted light-green arrow shows the phonon that would be required to mediate amplification in the backward direction, which is not supported by the acoustic dispersion relation. Thus, the elastic wave that mediates forward stimulated inter-modal Brillouin scattering does not mediate a backward scattering process.}
\label{fig:non}
\end{figure*}

An important feature of this Brillouin system is that the amplificaiton process is inherently nonreciprocal.  This property arises from the time modulation induced and sustained by the coherent phonon field. Alternatively, this phenomena can also be understood through the energy and phase-matching conditions required by this traveling-wave process (see Fig. \ref{fig:non}b-c).  The phonon field that mediates the Brillouin amplification cannot mediate coupling in the backward direction as a result of phase-matching. Moreover, the phonon frequency and wavevector that would be required for a backward process (between counter-propagating waves) is not supported by the acoustic dispersion relation.  These phase-matching and energy-conservation considerations are illustrated succinctly by the diagrams presented in Fig. \ref{fig:non}b-c.

We next detail the experimental approach and apparatus used to investigate the nonreciprocal nature of this resonantly enhanced Brillouin amplifier (data presented in Fig. 3c of the main text). As diagrammed in Fig. \ref{fig:non}a, these measurements involve the use of a fiber-optic switch, which allows us to rapidly change the propagation direction of the signal wave relative to the pump wave. Throughout the experiments, the pump wave is coupled into a counter-clockwise antisymmetric mode of the resonator that is separated from a symmetric cavity mode by the Brillouin frequency ($\omega_{\rm p}=\omega^m_2$, $\omega^m_2-\omega^n_1=\Omega_{\rm B}$). We then measure the signal wave transmission in the forward and backward directions as we sweep the signal-wave frequency $\omega_{\rm s}=\omega_{\rm p}-\Omega$ through the dual resonance condition ($\omega_{\rm s}=\omega^n_1=\omega_{\rm p}-\Omega_{\rm B}$).

\begin{table}[ht]
\begin{center}
\begin{tabular}{ |c|c|c| } 
\hline 
Optical properties & Description & Value\\
\hline
L & Length of device &1.57 cm  \\ 
$W_{\rm m}$ & Acoustic membrane width & 2.85 $\upmu$m  \\ 
$W_{\rm o}$ & Optical waveguide width & 1.5 $\upmu$m  \\ 
$\textup{FSR}_1$ & Free spectral range of symmetric mode & 5.02 GHz \\
$\textup{FSR}_2$ & Free spectral range of antisymmetric mode & 4.87 GHz\\
$\mu^2_{\rm A,1}$ &Coupling through coupler A into the symmetric mode &0.045-0.16  \\
$\mu^2_{\rm A,2}$ & Coupling through coupler A into the antisymmetric mode&0.62-0.71  \\
$\mu^2_{\rm B,1}$ & Coupling through coupler B into the symmetric mode&0.01-0.07  \\
$\mu^2_{\rm B,2}$ & Coupling through coupler B into the antisymmetric mode&0.007-0.016  \\
$\alpha_1$ & Symmetric spatial mode propagation loss&6.7 $\rm m^{-1}$  \\ 
$\alpha_2$ & Antisymmetric spatial mode propagation loss &20.6 $\rm m^{-1}$  \\ 
$v_{\rm g,2}$ & Symmetric spatial mode group velocity &7.91 $\times 10^7$ m/s  \\
$v_{\rm g,1}$ & Antisymmetric spatial mode group velocity&7.67 $\times 10^7$ m/s \\
$\Omega_{\rm B}$ & Brillouin frequency &$2\pi$ 5.95 GHz \\
$\Gamma$ & Acoustic dissipation rate&$2\pi$ 9 MHz \\
$G_{\rm B}$ & Brillouin gain coefficient &380 $\rm W^{-1} m^{-1}$\\
\hline
\end{tabular}
\end{center}
\caption{Experimental parameters of the nonreciprocal resonantly enhanced silicon Brillouin amplifier}
\label{tab:par}
\end{table}

\subsection{System parameters and data analysis}

In this section, we detail the experimental parameters of the system and how they are determined. All parameters in the system are corroborated by a self-consistent model of the amplifier dynamics.  We enumerate these parameters in Table \ref{tab:par}. The passive optical properties of the resonator system are obtained by fitting the transmission spectrum to a multi-spatial mode resonator model (for more details, see section B of this Appendix and Supplementary Materials of Ref. \cite{otterstrom2018silicon}). As the couplers A and B are wavelength dependent, we indicate an approximate range for each coupling coefficient.  The acoustic dissipation rate is determined by analyzing the Brillouin-gain bandwidths at low intracavity pump powers. The Brillouin-gain coefficient is extracted by fitting the mean-field model to our measurements (see Fig. 2c of the main text).  Altogether, these parameters are in agreement with those previously obtained in similar device geometries \cite{otterstrom2018silicon}. 

To determine the degree of net amplification (for Fig. 2 of the main text), we calibrate the signal-wave transmission through the resonant system by comparing it to the signal-wave transmission through a linear device on the same chip (placed 600 $\upmu$m away). Data for Fig. 2c-d of the main text were obtained by acquiring traces as the signal-wave detuning ($\Omega$) is swept through the Brillouin frequency ($\Omega_{\rm B}$)) while sweeping the the pump power and varying pump wavelength around the dual-resonance condition (i.e., $\omega_{\rm p}=\omega^m_2$, $\omega_{\rm p}-\Omega_{\rm B}=\omega^n_1$). The data in Fig. 2c-d represent the  traces obtained with near-zero detuning ($\omega_{\rm p}-\Omega_{\rm B}-\omega^n_1$).  Intracavity powers are estimated using the passive ring resonator parameters and accounting for the effects of nonlinear loss; additionally, this analysis is consistent with the power-dependent nature of the directional couplers.

\subsection{Device fabrication}

The Brillouin amplifier devices are fabricated on a single-crystal silicon-on-insulator chip (215 nm silicon, 3 $\upmu$m $\textup{SiO}_2$) using a two-step electron-beam lithography process (for additional details, see \cite{kittlaus2016large,kittlaus2017chip,otterstrom2018silicon}). Optical waveguides are defined through electron-beam lithography (using hydrogen silsesquioxane (HSQ) electron-beam resist), followed by development in MF-312 and an anisotropic $\textup{Cl}_2$ reative ion etch (RIE), which removes 80 nm of silicon. A subsequent lithography step is used to pattern slots with CSAR electron-beam resist.  After development in Xylenes, the remainder of the silicon is removed through another RIE etch, exposing the oxide underneath the slotted regions.  Finally, a wet-etch (49\% hydrofluoric acid) is used to remove the oxide undercladding to create a continuously suspended Brillouin-active waveguide.  The phonon membrane dimensions are 2.85 $\upmu$m $\times$ 135 nm, while the multimode optical ridge waveguide is 1.5 $\upmu$m $\times$ 215 nm \cite{kittlaus2017chip,otterstrom2018silicon}.

\twocolumngrid

\bibliography{cites}

\end{document}